\documentclass[preprint,showpacs,preprintnumbers,amsmath,amssymb,prb]{revtex4}

\usepackage{graphicx}
\usepackage{bm}

\begin{document}

\title{Optimization of the residence time of a Brownian particle in a spherical subdomain}

\author{O. B\'enichou}
\affiliation{Laboratoire de Physique Th\'eorique de la Mati\`ere Condens\'ee
(UMR 7600), case courrier 121, Universit\'e Paris 6, 4 Place Jussieu, 75255
Paris Cedex}

\author{R. Voituriez}
\affiliation{Laboratoire de Physique Th\'eorique de la Mati\`ere Condens\'ee
(UMR 7600), case courrier 121, Universit\'e Paris 6, 4 Place Jussieu, 75255
Paris Cedex}

\date{\today}

\begin{abstract}
In this communication, we show that the residence time of a Brownian particle, defined as the cumulative time spent in a given region
of space, can be optimized as a function of the diffusion coefficient.  We discuss the relevance of this effect to several schematic experimental situations, classified in the nature  -- random or deterministic -- both of the observation time and of the starting position of the Brownian particle.

\end{abstract}


\maketitle


Among the numerous features of stochastic processes, the time it takes a random walker
to  reach a target  -- the so-called first-passage time (FPT)  --  is a crucial quantity that governs a variety of physical systems
\cite{Kampen:1992,Redner:2001,Obenichou:2008,Condamin:2005db,Condamin:2007zl}.  Indeed, numerous real situations, ranging
from (sub)diffusion limited reactions  \cite{Rice:1985,S.B.Yuste:2002} to animals
searching for food \cite{Benichou:2005qd,Benichou:2006lq} can be rephrased as
first-passage problems. In all these situations, the FPT is a limiting quantity, whose optimization is a crucial issue.

\begin{figure}[htbp]
\centerline{\includegraphics[scale=0.8]{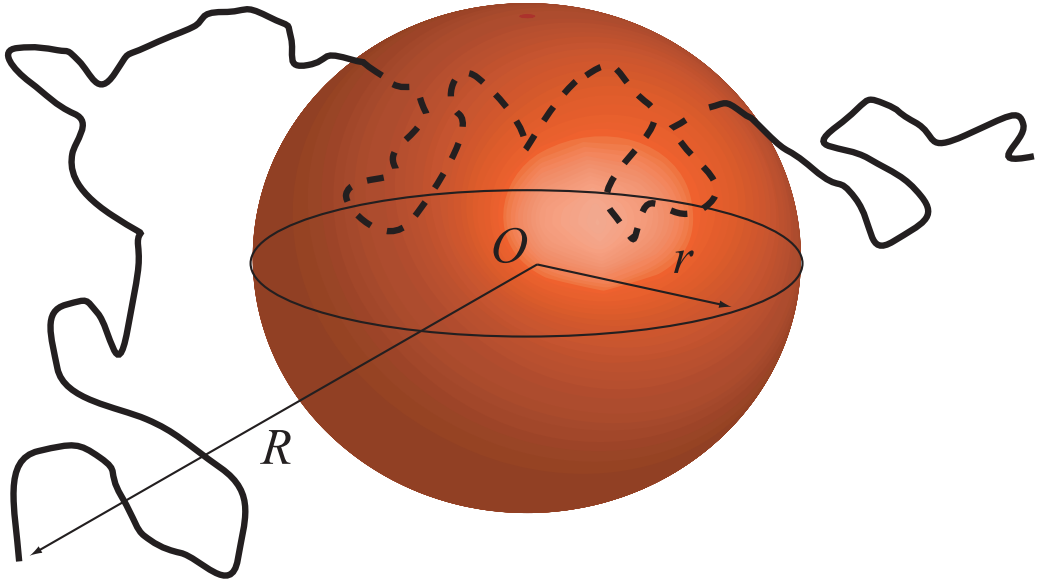}}
\caption{A Brownian particle crossing an interaction sphere. The residence time is the cumulative time spent by the particle inside the sphere.}
\label{def}
\end{figure}

Very often however, the reaction process is not infinitely efficient, so
that the relevant quantity which has to be optimized is  not the FPT
 to the target itself, but rather the residence time of the random walker in the
vicinity of the target  \cite{Agmon:1984}. The importance of this observable, defined as the cumulative time spent by the random walker in a given subdomain centered on the target up to a fixed observation time (see figure \ref{def}),  comes from the fact that it can be seen as a measure of the 
interaction time between the random walker and the target. The study of  the
statistics of this  general  quantity has been  a subject of
interest for long,   both for mathematicians \cite{darling,Aldous:1999,Hughes:1995a} and
physicists \cite{Wilemski:1973a,Agmon:1984,Godreche:2001,Majumdar:2002,Benichou:2003,Blanco:2003a,Benichou:2005a,Blanco:2006,Grebenkov:2007,Burov:2007,Condamin:2007rz,Condamin:2008}. As a matter of fact,  the residence time has
proven to be a key quantity  in various fields, ranging from
astrophysics \cite{Ferraro:2004}, transport in porous media \cite{Bouchaud:1990b}
 and 
diffusion limited reactions \cite{Wilemski:1973a,Agmon:1984,Benichou:2005hl}.

The question we address here is to determine how the residence time depends on transport
properties of the random walker, and more precisely on its diffusion coefficient. As we proceed to show, 
this dependence is non trivial, and in some situations it
is possible to tune the value of the diffusion coefficient in order to maximize the
residence time. 

To get an intuition of this effect, let us start with a simple 
analysis of the problem. Each time the random walker enters the interaction zone of typical size $r$, it spends inside
a typical time of order $r^2/D$
which is a decreasing function of the diffusion coefficient $D$. On the other hand, after an
observation time $t$, the number of times the interaction zone has been visited is an
increasing function of $D$, and behaves like
$\sqrt{Dt/r^2}$ in dimension 1, $\ln(Dt/r^2)$ in dimension 2 and like a constant in 
dimension $d\ge 3$, in the long time limit $t\to\infty$ \cite{Hughes:1995a}. For a walker starting 
from the exterior of the interaction zone, it is thus clear that both limits of small and large
diffusion coefficient $D$ lead to a small residence time, indicating the existence of a maximum as a function of $D$. However, it has
to be noted that the observation time has to be {\it finite} to make this optimization possible. Indeed, the mean residence
time $\mu$ goes to infinity with the observation time  in dimension $d\le 2$, and tends to a constant divided by $D$ 
in dimension $d\ge 3$, which is a decreasing function of $D$. 

With the exception of  \cite{refId} -- which only deals with  a one-dimensional situation in the context of DNA/protein interactions -- and of \cite{Golding:2006,Guigas:2008,Lomholt:2007,Eaves:2008,Zaid:2009} -- where it has been proposed that slow diffusion could be a way of enhancing the kinetics of imperfect reactions -- the dependance of $\mu$ with $D$ does not seem to have received a lot of attention
so far, probably for two reasons. First, in most of calculations
of residence times, the diffusion coefficient is taken fixed ($D=1/2$ in the mathematical
literature, $D=1$ in \cite{Berezhkovskii:1998}). Second,  the observation time is generally taken to be infinite
or at least large in explicit determinations of residence times (see for instance the Section III of \cite{Berezhkovskii:1998}),  which, as mentioned above,  does not permit to reveal the
non monotonic behavior we are looking for.

In the following, we focus on the generic case of a Brownian particle evolving in a 3--dimensional space, and discuss  three "experimental" situations, potentially relevant to different
problems of chemical reactivity : (i) the observation time is a deterministic variable ; (ii) the observation
time is a random variable ; (iii) both the observation time and the starting position
of the particle are random.

Denoting ${\bf R}$ the initial position of the particle and $r$ the radius of the target, assumed 
to be a sphere centered at the origin $O$, we study the pdf $P_{t}(T)$ of the residence time
$T$ after an observation time $t$. Following Kac \cite{darling}, we introduce the following trapping
problem : we assume that the Brownian particle disappears on the target
with a constant and uniform rate $k$. The survival probability $S(t)$ after an observation
time $t$ can be written in two different ways. First, it reads
\begin{equation}
S(t)=\int_{0}^t P_{t}(t')e^{-k t'}{\rm d}t',
\end{equation}
where $e^{-k t'}$ represents the survival probability at time $t'$ conditioned by the fact
that the particle has spent a time $t'$ is the reactive zone. Second, it is easy to see that $S$
satisfies the backward Fokker-Planck equation \cite{Redner:2001}:
\begin{equation}\label{surv}
\partial_{t}S(t)=[D\Delta_{{\bf R}}-kH(r-R)]S(t),
\end{equation}
where $H(x)=1$ if $x>0$ and $0$ otherwise,  the laplacian 
$\Delta_{{\bf R}}$ involves derivatives with respect to the starting point ${\bf R}$ and $R\equiv |{\bf R}|$.
 This partial 
differential equation has to be completed by the initial condition $S(t)\to 1$ if $t\to 0$ 
and the boundary condition $S(t)\to 1$ if $R\to \infty$. 
Denoting by ${\widehat S}(s)=\int_{0}^\infty e^{-s t}S(t){\rm d}t$ the Laplace transform 
of the survival probability and Laplace transforming Eq. (\ref{surv}), it is easily seen
that, if the starting point ${\bf R}$ is exterior to the sphere of radius $r$,
\begin{equation}
\label{1}
{\widehat S}(s)=\frac{1}{s}+AR^{1-d/2} K _{d/2 -1}\left(R\sqrt{\frac{s}{D}}
\right),
\end{equation}
while if ${\bf R}$ is interior
\begin{equation}
\label{2}
{\widehat S}(s)=\frac{1}{s+k}+BR^{1-d/2} I _{d/2 -1}\left(R\sqrt{\frac{s+k}{D}}
\right),
\end{equation}
where 
the constants $A$ and $B$ are given by continuity conditions of the survival
probability and of its first spatial derivative at $R=r$, $K_\nu$ and $I_\nu$ stand for modified Bessel functions and the space dimension is $d$.  Note that, in Eqs.(\ref{1}),(\ref{2}), only one of the two Bessel functions  $I _{d/2 -1}$ and  $K _{d/2 -1}$ generating the set of solutions appears, in order to fulfill boundary conditions when $R\to 0$ and $R\to \infty$.  In our case $d=3$,  
 it is explicitly found that, if the starting point is
  exterior to the sphere of radius $r$,
  \begin{eqnarray}
\label{laplaceS2}
{\widehat S}(s)&=&\frac{1}{s}+\frac{e^{-(R-r)\sqrt{s/D}}}{R}\times\nonumber\\
&\times&\left[\frac{k}{s(s+k)}\frac{\sinh(r\sqrt{(s+k)/D})-r\sqrt{(s+k)/D}\cosh(r\sqrt{(s+k)/D})}{\sqrt{s/D}\sinh(r\sqrt{(s+k)/D})+\sqrt{(s+k)/D}\cosh(r\sqrt{(s+k)/D})}\right],\nonumber\\
\end{eqnarray}
while if it is interior:
\begin{eqnarray}
\label{laplaceS1}
{\widehat S}(s)&=&\frac{1}{s+k}+\frac{\sinh(R\sqrt{(s+k)/D})}{R}\times\nonumber\\ 
&\times&\left[\frac{k(1+r\sqrt{s/D})}{p(p+k)}\frac{1}{\sqrt{s/D}\sinh(r\sqrt{(s+k)/D})+\sqrt{(s+k)/D}\cosh(r\sqrt{(s+k)/D})}\right].\nonumber\\
\end{eqnarray}

The Laplace transform  ${\widehat \mu}(s)$
of the mean residence time $\mu(t)$ is then obtained by expanding Eqs(\ref{laplaceS2})-(\ref{laplaceS1}) as a function of the parameter $k$, since ${\widehat S}(s)$ writes:
\begin{eqnarray}
{\widehat S}(s)&=&\int_0^\infty {\rm d} te^{-s t} \int_0^t {\rm d} t' e^{-k t'}P_t(t')=\frac{1}{s}- k  {\widehat \mu}(s) + o(k)
\end{eqnarray}
If the starting point is 
exterior to the sphere of radius $r$, the small $k$ expansion of Eq(\ref{laplaceS2}) leads to :
\begin{align}
\label{Laplace2}
{\widehat \mu}(s)=\frac{1}{s^2}\frac{e^{-R\sqrt{s/D}}}{R\sqrt{s/D}}[
{r\sqrt{s/D}\cosh(r\sqrt{s/D})-\sinh(r\sqrt{s/D})}],
\end{align}
while if the starting point is interior to the sphere of radius $r$, the small $k$ expansion of Eq(\ref{laplaceS1}) leads to :
\begin{eqnarray}
\label{Laplace1}
{\widehat \mu}(s)=\frac{1}{s^2}\left(1-\frac{(1+r\sqrt{s/D})
\sinh({R\sqrt{s/D}})}{R\sqrt{s/D}\exp(r\sqrt{s/D})}\right),
\end{eqnarray}

\begin{figure}[htbp]
\centerline{\includegraphics[scale=0.8]{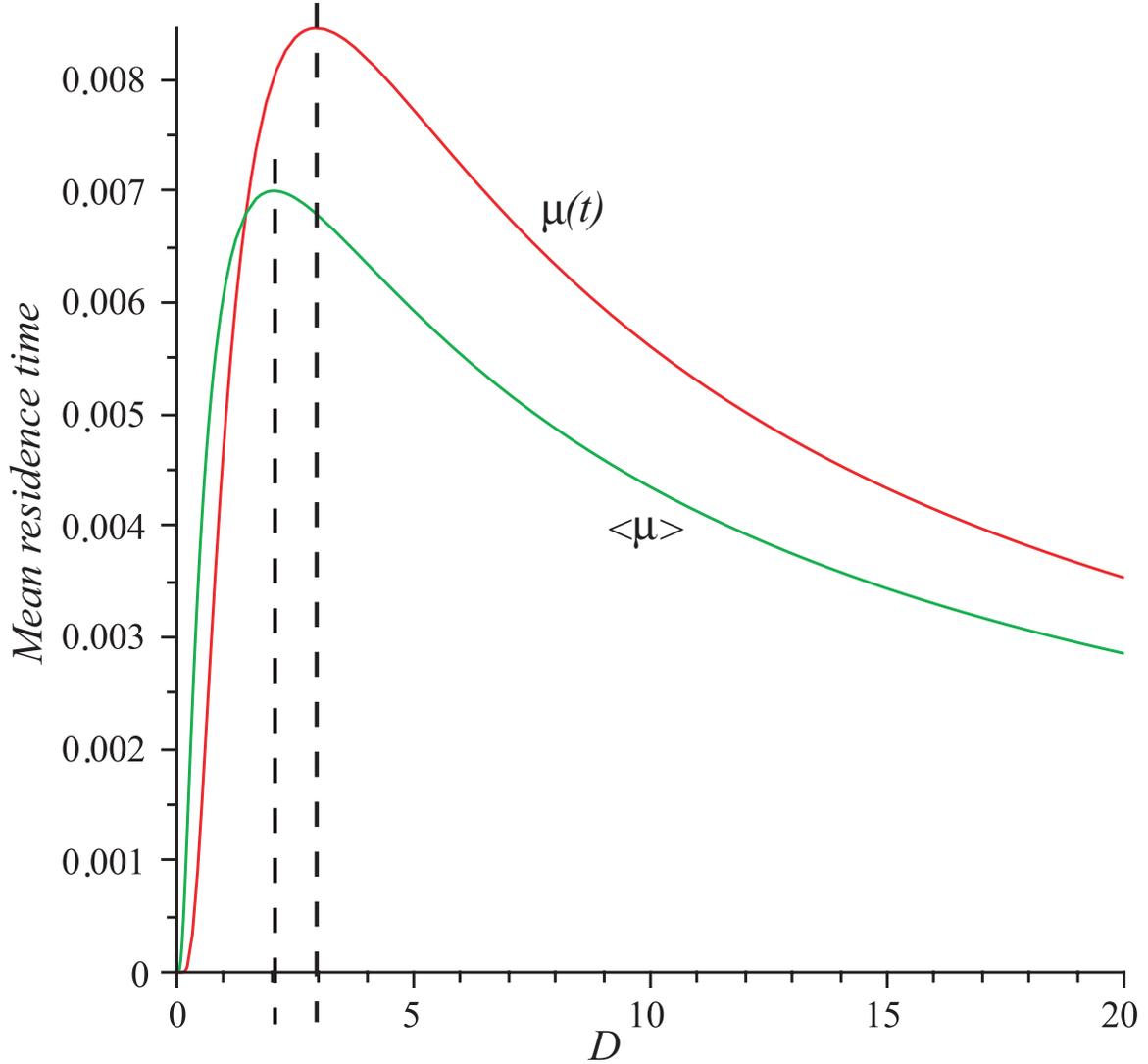}}
\caption{The mean residence time as a function of the diffusion coefficient for deterministic ($t=1$) and random observation times ($p=1$). In both cases, the Brownian particle starts outside of the sphere  ($R=3$, $r=1$). The optimal 
diffusion coefficient in the deterministic case (resp. random) is well approximated by the limiting expression $D_0\approx \alpha R^2/t$ (resp. $D_0\approx p R^2/4$) given in the text, even if here the condition of applicability $R\gg r$ is not
well satisfied.}
\label{optimisation}
\end{figure}

{\it Deterministic observation time.} In the first case of a deterministic observation time $t$, corresponding for instance to a chemical reaction designed to be terminated after a time $t$, it is actually possible
to Laplace invert the transform of the mean residence time. Using the two known Laplace inverts:
\begin{equation}
{\cal L}^{-1}(e^{-\sqrt{s}}/s^2)(t)=-\sqrt {{\frac {t}{\pi }}}{{\rm e}^{-1/(4\,{t})}}+\frac{1}{2}\,\left(1-{\rm erf}
 \left({\frac {1}{2\sqrt {t}}} \right)\right)  \left( 2\,t+1 \right) 
\end{equation}
\begin{equation}
{\cal L}^{-1}(e^{-\sqrt{s}}/s^{5/2})(t)=-\frac{1}{6}\,\left(1-{\rm erf} \left({\frac {1}{2\sqrt {t}}} \right)\right)  \left( 1+
6\,t \right) +\frac{1}{3}\,\sqrt {{\frac {t}{\pi }}}{{\rm e}^{-1/(4\,{t})}}
 \left( 4\,t+1 \right)
\end{equation}
where ${\cal L}^{-1}$ stands for the inverse Laplace operator and ${\rm erf}(x)$ is the error function, a direct Laplace inversion of  Eqs.(\ref{Laplace2}),(\ref{Laplace1})
gives the explicit expression
\begin{equation}
\label{deter1}
\mu(t)=\phi (r,R;t)-\phi (-r,R;t)+
\left\{\begin{array}{lll}
\displaystyle \frac{1}{2D}(r^2-R^2/3),  \;\;\;   {R<r}, \nonumber\\
\nonumber \\
\displaystyle \frac{r^3}{3RD},  \;\;\;   {R>r},
\end{array}
\right.
\end{equation}
where
\begin{eqnarray}
&&\phi(r,R;t)=\frac{2}{3}\sqrt{\frac{D}{\pi}}\frac{t^{3/2}}{R}
\left(1+\frac{(R+r)(R-2r)}{4Dt}\right)e^{-\frac{(R+r)^2}{4Dt}}\nonumber\\
&&+\frac{1}{2}{\rm erf}\left(
\frac{r+R}{2\sqrt{Dt}}\right)\left(t+\frac{R^2}{6D}-\frac{r^3}{3DR}
-\frac{r^2}{2D}\right).
\end{eqnarray}
Note that in the limit of large observation times $t\to \infty$, this expression gives back the known results of  \cite{Berezhkovskii:1998} :
$\displaystyle \mu(\infty)=\frac{3r^2-R^2}{6D}$ if $R<r$ (compare with Eq.(3.16a) of \cite{Berezhkovskii:1998}) and  $\displaystyle \mu(\infty)=\frac{r^3}{3 R D}$ if $R>r$ (compare with Eq.(3.11) of \cite{Berezhkovskii:1998}). Moreover, as expected, it is easily seen that, if $R>r$, $\mu(t)\to 0$ if $D\to 0$ or if $D\to \infty$. 
As a consequence, there exists a value of $D_0$ that optimizes the residence time (see figure \ref{optimisation}). 
Deriving the expressions of $\mu(t)$ with respect to $D$, this optimal value is easily seen to satisfy the following implicit equation:
\begin{eqnarray}
\label{derivee}
&& -2\,{{\rm e}^{-{
\frac { \left( R+r \right) ^{2}}{ 4 D_0  t}}}} \sqrt{ D_0
  t} \left( {R}^{2}-2\,{r}^{2}-Rr-2\,  D_0  t
 \right) +2\,{{\rm e}^{-{\frac { \left( -r+R \right) ^{2}}{4
  D_0 t}}}}  \sqrt{D_0  t} \left( {R}^{2}+Rr-2\,
  D_0  t-2\,{r}^{2} \right)\nonumber\\ 
  &&-
\sqrt {\pi }{{\rm erf}\left({\frac {R+r}{2\sqrt {  D_0  t}}}\right)}
 \left( R-2\,r \right)  \left( 
R+r \right) ^{2}+\sqrt {\pi }
{{\rm erf}\left({\frac {-r+R}{2\sqrt {  D_0  t}}}\right)}
 \left( 2\,r+R \right)  \left( -r+R \right) ^{2}=0
\end{eqnarray}
whose  solution can be discussed in limiting regimes. 
When $R\gg r$, we find that $D_0\sim \alpha R^2 /t$, where $\alpha\simeq 0.353$ is the implicit solution of the equation
\begin{equation}
 2\sqrt{\pi }\left[{\rm erf}\left(\frac{1}{2\sqrt{\alpha}}\right)-1\right]+\frac{e^{-1/(4\alpha)}}{\sqrt{\alpha}}=0,
\end{equation}
obtained  by replacing $D_0$ by  $\alpha R^2 /t$ in Eq.(\ref{derivee}) and taking the limit $R\gg r$.
In other words, the optimal diffusion coefficient $D_0$ does not depend of  the extension of the target $r$  in this regime. This is in strong contrast with the opposite regime
 $R\to r^+$,  which leads after a similar analysis to an  optimal diffusion coefficient
 $D_0\sim 3 r^2(R/r-1)/(2t)$,
which, in particular,  vanishes when $R\to r^+$ as could be expected from the qualitative analysis presented above.

{\it Random observation time.} We now consider another experimental situation, corresponding to a random
observation time, distributed according to an  exponential law with mean $1/p$.
This case corresponds for example to a situation where the particle has a finite lifetime, as in
many biological situations. Averaging over this random lifetime, the mean residence
time becomes:
\begin{eqnarray}
\langle \mu \rangle \equiv\int _{0}^\infty {\rm d}t p e^{-pt}\int_{0}^t {\rm d}t' 
 t'P_{t}(t')\equiv p {\widehat \mu}(s=p),
\end{eqnarray}
where $ {\widehat \mu}(s)$ is given by Eqs.(\ref{Laplace1})-(\ref{Laplace2}).
Once again, for $R>r$, the mean residence time can be shown to have an optimum as a 
function of $D$ (see figure \ref{optimisation}). Interestingly, 
the optimal diffusion coefficient $D_0$ satisfies the simple implicit equation
\begin{equation}
\label{implicite}
\tanh{x}=\frac{ax^2+x}{x^2+ax+1},
\end{equation}
where $x=r\sqrt{p/D_0}$ and $a=R/r$. 

For $a\gg1$ , the Lagrange inversion formula \cite{Wilf:1990} allows one to obtain explicitly the expansion of $x$ as powers of $1/a$, up to arbitrary order. The first terms are given by:
$ x=\frac{2}{a}+\frac{4}{5a^3}+...$.
In the opposite limit $a\to 1$, it is easy to deduce from Eq. (\ref{implicite}) that
$\displaystyle x\sim\frac{1}{\sqrt{a-1}}$. Finally, this gives again two regimes for the optimal diffusion coefficient: $D_0\sim p R^2/4$ when $R\gg r$ and  $D_0\sim p r^2(R/r-1)$ when $R\to r^+$, in close analogy with the two regimes found in the case of a deterministic observation time.

{\it Random observation time and random initial position.} In the last situation, we assume that both the observation time and the initial position are random variables. More precisely, the observation 
time is again assumed to be distributed according to an exponential law with mean $1/p$, while 
the initial position is assumed to be uniformly distributed in an annulus region centered on 0, of inner radius $b$ and outer radius $c$. This mimics  for example a case of chemical reactions 
where the position of the source of reactants cannot be accurately determined. In this case
of an annulus region, it is not clear {\it a priori} if the mean residence time is still non trivially optimizable with respect to $D$. Indeed, if the initial position is within the reactive sphere, the best diffusion coefficient is
simply $D_0=0$. On the contrary, if the annulus region is entirely exterior to the reactive sphere, a non trivial optimal $D_0$ is expected from previous results. This raises the question of the intermediate case where the annulus partially covers the reactive sphere.

In the first case of an annulus region exterior to the target zone ($b>r$), we find
\begin{eqnarray}
 &&\overline{\langle \mu \rangle}=\frac{3rl}{p(c^3-b^3)}\left[\frac{\sinh(r/l)}{r/l}-\cosh(r/l)\right]\nonumber\\
&\times& \left[\left(c+l\right)e^{-c/l}-\left(b+l\right)e^{-b/l}\right],
\end{eqnarray}
where $l\equiv \sqrt{D/p}$.
This quantity is easily shown to tend towards 0 when $D\to \infty$ while
\begin{equation}
 \overline{\langle \mu \rangle}\sim_{D\to 0} \frac{3rbl}{2p(c^3-b^3)} e^{-(b-r)/l} >0.
\end{equation}
This proves that $\overline{\langle \mu \rangle}$ admits an optimum diffusion 
coefficient, as  expected in view of the previous results.

In the opposite situation of a annulus region such that $0<b<r<c$, we get
 \begin{align}
 \label{eqnn}
&\overline{\langle \mu \rangle}-\frac{1}{p}\frac{r^3-b^3}{c^3-b^3}= \frac{3l}{p(c^3-b^3)}\times \nonumber\\
&\left[
e^{-r/l}\left(r+l\right)\left(b\cosh \left(b/l\right) - l\sinh \left(b/l\right)\right)\right.\nonumber\\
&\left.
-e^{-c/l}\left(c+l\right)\left(r\cosh \left(r/l\right) - l\sinh \left(r/l\right)\right)
\right].
\end{align}
The local behavior of this quantity when $D\to 0$ actually depends on the relative order of the two distances $r-b$ and $c-r$.
If $r-b<c-r$, 
\begin{equation}
\overline{\langle \mu \rangle}-\frac{1}{p}\frac{r^3-b^3}{c^3-b^3}\sim  \frac{3rbl}{2p(c^3-b^3)} e^{-(r-b)/l}>0,
\end{equation}
showing that there is an optimal  diffusion coefficient.

Remarkably, if $r-b>c-r$, 
\begin{equation}
\overline{\langle \mu \rangle}-\frac{1}{p}\frac{r^3-b^3}{c^3-b^3}\sim  -\frac{3}{p(c^3-b^3)}\sqrt{\frac{D}{p}} \frac{rc}{2}e^{-(c-r)\sqrt{p/D}}<0,
\end{equation}
and this time there is no possible optimization.

The last case of an annulus region of inner radius equal to zero (i.e. $b=0$), i.e. a sphere of radius $c$,
deserves some specific attention. In this case, Eq. (\ref{eqnn}) is still valid but, quite surprisingly, no optimization 
with respect to the diffusion coefficient is possible. This monotonic decrease 
is compatible with the local  behavior ($D\to0$):
\begin{equation}
\overline{\langle \mu \rangle}-\frac{1}{p}\frac{r^3}{c^3}\sim  -\frac{3r}{2pc^2}\sqrt{\frac{D}{p}} e^{-(c-r)\sqrt{p/D}}<0.
\end{equation}
We stress that this result holds for arbitrary values of the radius $c$.

In conclusion, we have proposed a very simple mechanism that allows one to optimize the residence time of diffusing molecules with respect to the diffusion coefficient. We have discussed 
the relevance of this effect to several schematic experimental situations, classified in the nature  -- random or deterministic -- both of the observation time and of the starting position. We believe that this optimization of the residence time discussed in the case of the standard Brownian motion is robust (see in particular Supplementary Material for a discussion of a discrete space version of the model presented here) and is generalizable to more complex transport processes.


\newpage

{\bf Figure captions:}

{\bf Fig1:} A Brownian particle crossing an interaction sphere. The residence time is the cumulative time spent by the particle inside the sphere.

\vspace {1cm}

{\bf Fig2:} The mean residence time as a function of the diffusion coefficient for deterministic ($t=1$) and random observation times ($p=1$). In both cases, the Brownian particle starts outside of the sphere  ($R=3$, $r=1$). The optimal 
diffusion coefficient in the deterministic case (resp. random) is well approximated by the limiting expression $D_0\approx \alpha R^2/t$ (resp. $D_0\approx p R^2/4$) given in the text, even if here the condition of applicability $R\gg r$ is not
well satisfied.

\newpage

\newpage

\end{document}